\documentclass[journal]{IEEEtran}
\usepackage{xcolor,soul,framed} 
\usepackage{graphicx}
\usepackage{float}
\usepackage{graphicx}
\usepackage{adjustbox}
\usepackage{color}
\usepackage{algorithm}
\usepackage{algorithmic}
\usepackage{amsfonts}
\usepackage{mathtools}
\usepackage{tabularx}
\usepackage{subfigure}
\colorlet{shadecolor}{yellow}

\usepackage{array}
\usepackage{amsmath}
\DeclareMathOperator*{\argmax}{argmax} 
 
\usepackage{mdwmath}
\usepackage{mdwtab}
\usepackage{eqparbox}
\usepackage{url}

\begin{document}
    \title{Human-Centric Resource Allocation for the Metaverse with Multi-Access Edge Computing}
\author{Zijian Long, Haiwei Dong,~\IEEEmembership{Senior Member,~IEEE}, and Abdulmotaleb El Saddik,~\IEEEmembership{Fellow,~IEEE}
\thanks{Manuscript received 2023. The work of Zijian Long was supported in part by the China Scholarship Council (No. 202107970008). (Corresponding author: Haiwei Dong.)}
\thanks{Zijian Long and Abdulmotaleb El Saddik are with the Multimedia Communications Research Laboratory (MCRLab), the School of Electrical Engineering and Computer Science, University of Ottawa, Ottawa, ON K1N 6N5, Canada (e-mail: \{zlong038, elsaddik\}@uOttawa.ca).}
\thanks{Haiwei Dong is with Ottawa Research Center, Huawei Technologies Canada, Ottawa, ON K2K 3J1, Canada (e-mail: haiwei.dong@ieee.org).}
\thanks{Copyright (c) 20xx IEEE. Personal use of this material is permitted. However, permission to use this material for any other purposes must be obtained from the IEEE by sending a request to pubs-permissions@ieee.org.}}
\markboth{IEEE Internet of Things Journal}%
{Shell \MakeLowercase{\textit{et al.}}: A Sample Article Using IEEEtran.cls for IEEE Journals}

\IEEEpubid{0000--0000/00\$00.00~\copyright~2021 IEEE}

\maketitle


\maketitle

\begin{abstract}
Multi-access edge computing (MEC) is a promising solution to the computation-intensive, low-latency rendering tasks of the metaverse. However, how to optimally allocate limited communication and computation resources at the edge to a large number of users in the metaverse is quite challenging. In this paper, we propose an adaptive edge resource allocation method based on multi-agent soft actor-critic with graph convolutional networks (SAC-GCN). Specifically, SAC-GCN models the multi-user metaverse environment as a graph where each agent is denoted by a node. Each agent learns the interplay between agents by graph convolutional networks with self-attention mechanism to further determine the resource usage for one user in the metaverse. The effectiveness of SAC-GCN is demonstrated through the analysis of user experience, balance of resource allocation, and resource utilization rate by taking a virtual city park metaverse as an example. Experimental results indicate that SAC-GCN outperforms other resource allocation methods in improving overall user experience, balancing resource allocation, and increasing resource utilization rate by at least 27\%, 11\%, and 8\%, respectively.
\end{abstract}

\begin{IEEEkeywords}
Extended reality, multi-agent reinforcement learning, attention mechanism, graph convolutional network.
\end{IEEEkeywords}

%
\IEEEpeerreviewmaketitle



\section{Introduction} \label{introduction}
\IEEEPARstart{T}{he} metaverse, regarded as the next generation of the Internet, has gained a lot of attention from both academia and industry. It is commonly defined as a set of virtual worlds in which people can work, play, and socialize through their respective avatars \cite{wang2022survey}. It integrates the most cutting-edge technologies, such as cloud/edge computing, artificial intelligence, eXtended reality (XR), digital twins, and blockchains \cite{el2018digital}. The current metaverse systems can roughly be classified into two categories: 1) multiplayer online games: Minecraft allows players using their avatars to explore, interact with, and modify a dynamically-generated 3D world made of blocks \cite{guss2019minerl}. Roblox, a 3D sandbox game, offers a programmable environment for players to design their worlds which they can share with others \cite{rospigliosi2022metaverse}. 2) social activity oriented metaverse systems: Meta created Horizon World, a popular virtual reality (VR) platform, to enable users to move around in a variety of worlds that host events, games, and social interactions \cite{gauci2018horizon}. Baidu also built Xirang where users interact and socialize with others using avatars in a virtual planet \cite{xu2022full}.

However, there are still many challenging problems remained to achieve the future metaverse where millions of users' virtual avatars live in a set of virtual worlds connecting closely to the physical world. For example, the metaverse requires a large amount of computation resources to render 3D virtual worlds in a seamless manner. Due to the limited computation resources of XR headsets, the computation-intensive rendering task cannot be solely conducted on them \cite{liu2018cutting}. Though remote rendering by powerful cloud servers can solve the insufficient computation resources problem, the metaverse users expect ultra-low latency of 20ms-30ms which cannot be satisfied by the current cloud-based infrastructures \cite{tang2022roadmap, metaverse_communication}. Multi-access edge computing (MEC) \cite{taleb2017multi}, which places powerful servers close to users, is a promising method to meet the high requirements of the metaverse for both communication and computation resources \cite{georgakopoulos2016internet}. 

\begin{figure*}[htbp]
\centering
\includegraphics[width=0.95\textwidth]{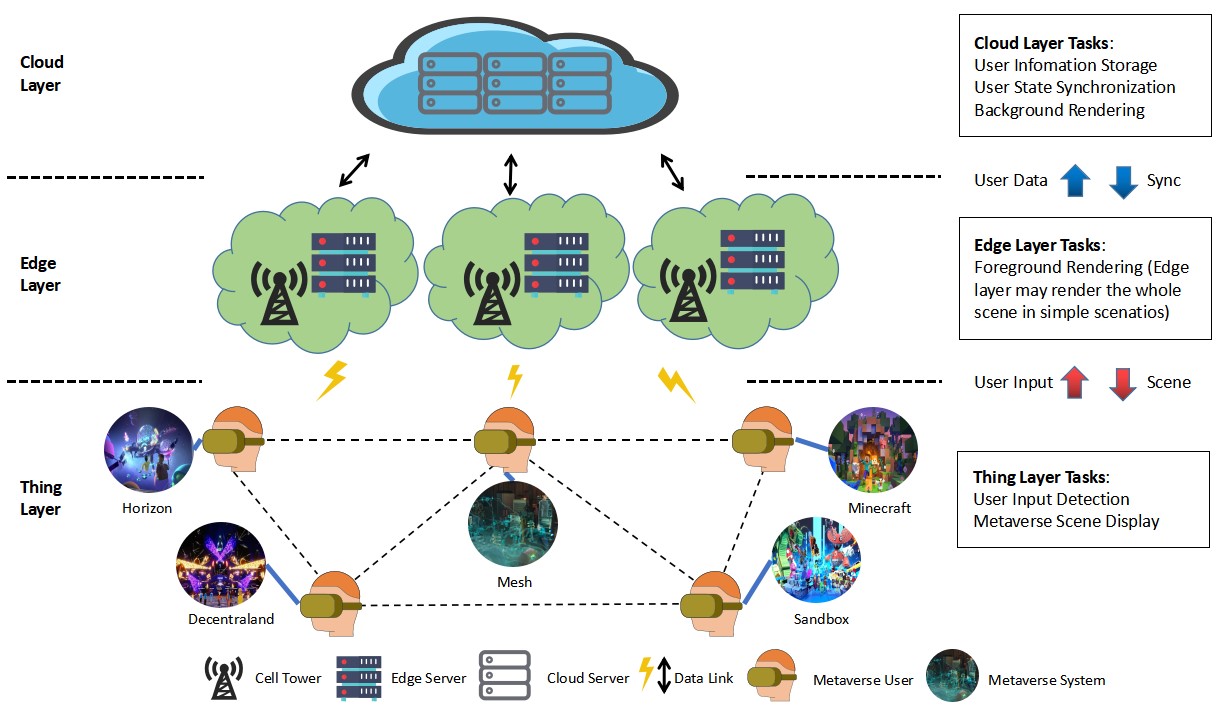}
\caption{The thing-edge-cloud collaborative architecture of MEC to provide multiple users with remotely rendered metaverse scenes. To enable human-centric resource allocation, agents that represent users are denoted as nodes and their interactions are described by dashed lines (explained in Section \ref{masac}).} 
\label{cloud}
\end{figure*}

In our vision, the thing-edge-cloud collaborative architecture of MEC is the key to enabling the metaverse, as shown in Fig. \ref{cloud}, which comprises three layers: the thing layer, the edge layer, and the cloud layer. In the thing layer, multiple users wear XR headsets to experience diverse 3D virtual worlds in the metaverse, such as Decentraland, Horizon, and Minecraft \cite{9984845}. XR headsets measure users' movement and transmit the data to the edge layer via the mice uplink flow. Users then receive rendered scenes of the metaverse by the large elephant downlink flow \cite{elbamby2018toward}. The edge layer is an intermediate layer between the thing layer and the cloud layer where edge servers are directly connected to cell towers \cite{luo2021resource}. 
\IEEEpubidadjcol
The edge servers mainly execute foreground rendering which has fewer requirements for graphical details but more requirements for stringent latency. Note that the edge layer may render both foreground and background in simple scenarios. The edge layer connects the thing layer and the cloud layer by transmitting user data from the thing layer to the cloud layer and delivering synchronization information from the cloud layer to the thing layer. The cloud layer consists of the most powerful computing and storage units in the data center. Even though the edge servers are capable of processing data rapidly, they still require the cloud layer to handle computation-intensive latency-tolerant operations, such as user information storage, user state synchronization, and background rendering \cite{ravindra2017mathbb, wu2020cloud}. 

Metaverse users can benefit from the thing-edge-cloud collaborative framework. However, how to adaptively allocate edge resources to multiple users in the metaverse has many challenges due to the following reasons. First, users' requests for resources change with their movements. A fast movement requires more computation resources to generate the responding scenes and more network resources to deliver the packets of that scenes. Second, the resource conditions are difficult to be predicted due to variable CPU utilization rates, network bandwidths, delays, and jitters. Third, the target of resource allocation is to improve the overall quality of experience (QoE) of users while keeping the balance of resource allocation among users. In the case of limited resources, if some users move faster they are provided more resources, which is determined by the QoE definition. Meanwhile, the balance of resource allocation would also be considered together for fairness. The trade-off between the two targets makes obtaining the optimal resource allocation strategy even more challenging.

In this paper, we address the above challenges in a human-centric way. First, the main target of our proposed method is to improve human experience which is defined by four human-related factors: user-received scene quality, frame choppiness, latency, and metaverse frame instability (More details will be given in Section \ref{problem}). Previous user studies have proven that these factors play key roles in determining human experience in the metaverse \cite{xu2022wireless, cheng2022we}. Second, to treat each metaverse user fairly, we add balance factor of resource allocation in the reward function. Moreover, our RL-based method is trained and tested with user data from a real metaverse resource allocation system rather than simulated data. The main contributions of our work are as follows:

\begin{itemize}
\item We propose a thing-edge-cloud collaborative framework of MEC to enable the metaverse at the network edge and formulate the problem of edge resource allocation as a decentralized partially observable Markov decision process (Dec-POMDP). 
\item To solve the Dec-POMDP problem, we propose a method based on multi-agent soft actor-critic with graph convolutional networks (SAC-GCN). In SAC-GCN, each agent adaptively determines the usage of communication and computation resources for one user in the metaverse. The former is presented as the bit rate of the rendered metaverse scenes, while the latter is described as the usage percentage of CPU.
\item To the best of our knowledge, this is the first time that combines SAC with graph convolutional networks to deal with the highly dynamic metaverse environment where agents keep moving and their neighbors change quickly. The environment is modeled as a graph where each agent is represented by a node. Edges are also added to neighboring nodes to indicate their mutual influence. Moreover, graph convolutional networks and self-attention mechanism are proposed to promote better cooperation between agents. 
\item We propose a smart edge resource allocation system for the metaverse where multiple users can access the metaverse wearing untethered VR headsets with adaptive communication and computation resources based on their requests. We also compare the performance of SAC-GCN with other resource allocation methods using a virtual city park as a case study. Experimental results indicate that SAC-GCN outperforms other resource allocation methods in improving overall user experience, balancing resource allocation, and increasing resource utilization rate by at least 27\%, 11\%, and 8\%, respectively.
\end{itemize}

The remainder of this paper is organized as follows. Section \ref{related} reviews the related work. Section \ref{problem} formulates the resource allocation as a Dec-POMDP and defines state space, action space, and reward in our considered scenario. Section \ref{masac} describes the proposed SAC-GCN. Section \ref{case} demonstrates the architecture of the resource allocation system and the performance of SAC-GCN using a virtual city park in the metaverse as an example. Experimental results are analyzed in Section \ref{analysis} followed by a conclusion in Section \ref{conclusion}.

\section{Related Work} \label{related}
In this section, we first briefly discuss the relationship between IoT and the metaverse. We then review the resource allocation problem in MEC as an enabler technology for the metaverse at the edge. We also discuss the state-of-the-art multi-agent deep reinforcement learning (MADRL) methods.
\subsection{Internet of Things and the Metaverse}
Internet of things (IoT) plays an important role in the architecture of the metaverse. First of all, the synchronization of data between the real world and the virtual world is essentially fundamental for the metaverse ecosystem \cite{cai2022compute}. Due to the recent rapid development of IoT, a large number of IoT sensors are being deployed to collect data from the real world which could be further shared with the virtual world. Han et al. proposed a dynamic hierarchical framework to synchronize metaverse with the collected data from a group of IoT devices based on optimal control theory \cite{han2022dynamic}. Secondly, IoT devices, such as VR and Mixed Reality (MR) headsets, are the most popular interfaces for users to interact with the metaverse \cite{han2022comic}. These IoT devices detect users' movements and recognize their gestures and voices to enable various modalities of interactions in the metaverse. For example, by wearing HoloLens 2 \cite{ungureanu2020hololens}, a MR headset manufactured by Microsoft, user is able to collaborate on persistent 3D objects with eye contact, facial expressions, and gestures \cite{dong2018real}.
\subsection{Resource Allocation for MEC}
MEC, as an emerging network paradigm, can provide extensive computation resources and reduce service time as well. However, intelligent resource allocation schemes are needed when dealing with the computation-intensive and latency-sensitive metaverse. Inspired by the success of deep reinforcement learning methods in sequential decision making, researchers have applied them to the edge resource allocation problem. Wang et al. proposed DRLRA to smartly allocate network resources and computing resources based on Deep Q-Network (DQN) \cite{wang2019smart}. The proposed method is deployed on a software-defined networking (SDN) \cite{ren2018novel} controller to collect the overview of the MEC environment, aiming to reduce the service time and balance resource consumption across the MEC servers. Liu et al. jointly considered the task offloading problem and resource allocation at the edge and proposed a multi-agent DQN-based framework to reduce system costs \cite{liu2020multi}. It is noted that edge resource allocation for the metaverse has bigger challenges due to its high requirements for the computation-intensive rendering and ultra-low latency streaming especially for a large number of users. 

\subsection{Multi-Agent Deep Reinforcement Learning}
MADRL algorithms are designed for the complex systems where multiple agents operate in a fully/partially shared-information environment. Each agent learns to make its own decisions by interacting with the environment and other agents at the same time. They are considered to be one of the promising solutions to the NP-hard problems. However, due to the non-stationarity of the environment, the main challenge in MADRL, it is quite difficult to make multiple agents learn and collaborate. Lowe et al. proposed a centralized training and decentralized execution framework for the multi-agent version of Deep Deterministic Policy Gradient (DDPG) \cite{lowe2017multi}. They employed a fully observable critic for each agent to deal with the global information during the training process. Jiang et al. modeled the multi-agent environment as a graph and employed graph convolutional networks to facilitate communication between nearby agents \cite{jiang2018graph}. Yang et al. proposed the mean field reinforcement learning method to solve the ``curse of dimensionality" caused by a large number of agents \cite{yang2018mean}. 

\section{Problem Formulation} \label{problem}
In the MEC-enabled metaverse, multiple users share the communication resources of the cell towers and the computation resources of the edge servers. How to intelligently allocate the resources to users can be seen as a multi-agent game where each agent determines how many resources a user should take. In general, multi-agent games can be placed into three groups: fully cooperative, fully competitive, and a mix of the two, depending on the types of settings they address \cite{yu2021surprising}. We set all the agents with a shared target which is to improve the overall QoE while balancing the resources allocated to each user. Therefore, all the agents collaborate to optimize this long-term target in a fully cooperative manner. As each agent only observes part of the environment, and all the agents have the same target, the problem can further be formulated as a Dec-POMDP, which is a special case of a partially observable Markov game (POMG) designed for cooperative interaction.


A Dec-POMDP can be defined as a tuple $(\mathcal{S}, \mathcal{A}, P, R, \mathcal{O},\\ \mathcal{N}, \Omega, \gamma)$ \cite{ross2008online}. At each time step $t$, the global state of the environment is denoted by $\boldsymbol{s}_t \in \mathcal{S}$. Each agent $i \in \mathcal{N} = \{1, 2, ..., N\}$ obtains its individual observation $o^i_t \in \mathcal{O}^i$ according to the observation function $\Omega(o^i_{t}|s_{t}): S \rightarrow O^i$. The agent then chooses an action $a^i_t \in \mathcal{A}^i$ based on its policy $\pi^i(a^i_t|o^i_t): \mathcal{O}^i \times \mathcal{A}^i \rightarrow [0, 1]$. All the actions selected by the agents at time step $t$ form a joint action $\boldsymbol{a}_t\ \in \mathcal{A} := \mathcal{A}^1 \times \mathcal{A}^2 \times ... \times \mathcal{A}^N$. The transition probability $P(\boldsymbol{s}_{t+1}|\boldsymbol{s}_t, \boldsymbol{a}_t): \mathcal{S} \times \mathcal{A} \times \mathcal{S} \rightarrow [0, 1]$ denotes the probability from the current environment state $\boldsymbol{s}_t$ to the next environment state $\boldsymbol{s}_{t+1}$ after the joint action $\boldsymbol{a}_t$ is executed. The reward function $R(\boldsymbol{s}_t, \boldsymbol{a}_t): \mathcal{S} \times \mathcal{A} \rightarrow \mathbb{R}$ describes the shared reward from the environment given the agents' joint action $\boldsymbol{a}_t$, where $\mathbb{R}$ represents the set of real numbers. $\gamma \in (0, 1]$ is a discount factor which determines how much the agents value the rewards in the future compared to those in the current state. In the following subsections, we define state space, action space, and reward for the metaverse resource allocation problem.

\subsection{State and Observation Space}
The observation of the environment for each agent includes the network condition $N^i_t$ and CPU performance $C^i_t$. The network condition can be represented by a six-parameter tuple: $N^i_t = (x^i_t, y^i_t, l^i_t, j^i_t, p^i_t, n^i_t)$, where $x^i_t$ is the last selected target bit rate; $y^i_t$ is the actually received bit rate; $l^i_t$ is the average round trip latency; $j^i_t$ is the network jitter; $p^i_t$ is the number of lost packets; and $n^i_t$ is the number of negative acknowledgment messages. The CPU performance is denoted by $C^i_t = (z^i_t, u^i_t, e^i_t, d^i_t)$, where $z^i_t$ is the last chosen number to limit CPU usage for agent $i$; $u^i_t$ is the percentage of overall available CPU; $e^i_t$ is the rendered frame rate at the edge server; and $d^i_t$ is the average delay for rendering a frame. Thus, the local observation is represented as the combination of these two sets of parameters
\begin{equation}
\begin{split}
o^i_t & = (N^i_t, C^i_t) \\ & = (x^i_t, y^i_t, l^i_t, j^i_t, p^i_t, n^i_t, y^i_t, u^i_t, e^i_t, d^i_t)
\label{local obs}
\end{split}
\end{equation}
We assume that the global state can be obtained by collecting all the agents' local observations. Therefore, the global state is defined as $\boldsymbol{s}_t = (o^1_t, o^2_t, ..., o^N_t) \in \mathcal{S}$, where $\mathcal{S}$ is the global state space. 
\subsection{Action Space}
An agent in our considered scenario executes actions to determine how many communication and computation resources a user is supposed to take. For communication resources, the options that an agent can choose from for the transmitting bit rate of the rendered frames are in the range $[b_{min}, b_{max}]$ Mbps. An agent's consumption of computation resources is controlled by throttling the CPU usage of the process that the agent targets. The options for the throttling usage percentage are in the range $[l_{min}, l_{max}]$. We define the action of agent $i$ as a tuple: $a^i_t = (a^{i, m}_t, a^{i, p}_t)$, where $a^{i, m}_t \in [b_{min}, b_{max}]$ and $a^{i, p}_t \in [l_{min}, l_{max}]$ represent the actions for communication and computation resources, respectively. Note that $a^{i, m}_t$ and $a^{i, p}_t$ both are positive numbers which means that the action space is a two-dimension continuous action space. Similar to the global state, the joint action is the combination of actions of all agents: $\boldsymbol{a}_t = (a^1_t, a^2_t, ..., a^N_t) \in \mathcal{A}$, where $\mathcal{A}$ is the action space.
\subsection{Reward Design} \label{user qoe}
Dec-POMDPs utilize shared rewards to evaluate the performance of joint actions and to guide agents to make better decisions in terms of long-term cumulative reward. When designing the reward function for $N$ users in the metaverse, three important factors that need to be considered are overall quality of experience (QoE) \cite{9754241}, balance of communication resource allocation, and balance of computation resource allocation. The last two factors represent balance of resource allocation in the metaverse.
\subsubsection{Overall QoE} 
We propose a time-step based QoE model defined as follows
\begin{equation}
\begin{split}
 QoE_t = & \underbrace{\alpha\sum_{t=1}^{T}q(y^i_t)}_{Metaverse\ Scene\ Quality}-\underbrace{\beta \sum_{t=1}^{T}|f^i_t - f^i_{target}|}_{Choppiness\ Penalty} \\ & -\underbrace{\gamma \sum_{t=1}^{T}p(l^i_t)}_{Latency\ Penalty}  -\underbrace{\delta \sum_{t=1}^{T-1}|q(y^i_{t+1})-q(y^i_t)|}_{Instability\ Penalty}
\end{split}
\label{qoe}
\end{equation}
For the time step $t$, $\alpha q(y^i_t)$ represents the overall level of satisfaction with the metaverse scene quality of agent $i$, where $y^i_t$ is the average received bit rate and $\alpha$ is the satisfaction level factor. The higher the bit rate, the better the scene quality and the more enjoyable the viewing and interaction of metaverse would be. Other items in the equation are used to penalize the negative impact of other major factors. $\beta |f^i_t - f^i_{target}|$ donates the choppiness penalty generated by lost frames where $f^i_t$ and $f^i_{target}$ are the received frame rate and the target frame rate and $\beta$ is the choppiness penalty factor. $\gamma p(l^i_t)$ is used to penalize the turnaround latency with $l^i_t$ being the average latency within time step $t$ and $\gamma$ being the latency penalty factor. The instability penalty for the changes in scene quality from time step $t$ to time step $t+1$ is represented by $\delta |q(y^i_{t+1})-q(y^i_t)|$ with $\delta$ as the instability factor. Because the marginal improvement in perceived quality decreases at higher bit rates, we use the logarithmic function to represent $q(y^i_t)$, where $q(y^i_t)=log(y^i_t/y^i_{min})$ and $y^i_{min}$ is the minimum value of the bit rate of all the time. A user's satisfaction level decreases more as the total latency increases, and higher latency significantly reduces QoE with sickness. Therefore, we use an exponential function to denote $p(l^i_t)=e^{l^i_t/l^i_{min}}$, where $l^i_{min}$ is the minimum value of the latency. The overall user experience at time step $t$ is represented by the sum of the QoE score $QoE^i_t$ for each agent $i$

\begin{equation}
Q_t = \sum_{i=1}^{N}QoE^i_t 
\label{a_qoe}
\end{equation}
\subsubsection{Balance of Communication Resource Allocation}
The obtained communication resources for agent $i$ at time step $t$ are presented as the bit rate of the metaverse scene $y^i_t$. We use variance of the bit rates to denote balance of communication resource allocation
\begin{equation}
V^{comm}_t = \frac{\sum_{i=1}^{N}(y^i_t - \frac{\sum_{i=1}^{N}y^i_t}{N})^2 }{N - 1} 
\label{v_comm}
\end{equation}
\subsubsection{Balance of Computation Resource Allocation}
The obtained computation resources for agent $i$ at time step $t$ are denoted as the percentage of CPU usage $u^i_t$. We use variance of the CPU usage to denote balance of computation resource allocation
\begin{equation}
V^{comp}_t = \frac{\sum_{i=1}^{N}(u^i_t - \frac{\sum_{i=1}^{N}u^i_t}{N})^2 }{N - 1} 
\label{v_comp}
\end{equation}
Finally, the time-step-based reward function is designed to deal with the trade-off between overall user experience and balance of resource allocation, denoted as $r_t = w_1 \cdot Q_t + w_2 \cdot V^{comm}_t + w_3 \cdot V^{comp}_t$. 

Given the definitions of state space, action space, and reward in our Dec-POMDP model, the target is to find the optimal joint policy $\pi^*$ which guides the agents to execute a joint action $\boldsymbol{a}_t$ at any state $\boldsymbol{s}_t$ to maximize the expected discounted return $G_t$, defined as
\begin{equation}
\begin{split}
\pi^* &= \argmax_{\pi}\ \mathbb {E}_\pi [G_t] = \argmax_{\pi}\ \mathbb {E}_\pi [\sum^{\infty}_{k=0} \gamma^kr_{t+k+1} ]
\label{policy}
\end{split}
\end{equation}
where $\gamma^k$ is the discount factor and $r_{t+k+1}$ is the reward for the time step $t+k+1$. The value of the global state $\boldsymbol{s}$ can be calculated by the state value function
\begin{equation}
V^\pi(\boldsymbol{s}) = \mathbb {E}_{\pi} [G_t|\boldsymbol{s}_t = \boldsymbol{s}]
\label{state value}
\end{equation}
Here $\pi$ is the current policy. The global state-action value function (Q-function) is defined as
\begin{equation}
\begin{split}
Q^\pi(\boldsymbol{s}, \boldsymbol{a}) & = \mathbb {E}_{\pi} [G_t|\boldsymbol{s}_t = \boldsymbol{s}, \boldsymbol{a}_t = \boldsymbol{a}] 
\label{state action value}
\end{split}
\end{equation}
Based on the above formulation, our Dec-POMDP problem can be solved by either model-based reinforcement learning methods or model-free reinforcement learning methods. The model-based methods are based on the state transition probability $P(\boldsymbol{s}_{t+1}|\boldsymbol{s}_t, \boldsymbol{a}_t)$ which denotes the probability from one state $\boldsymbol{s}_t$ to the next state $\boldsymbol{s}_{t+1}$ with action $\boldsymbol{a}_t$. Due to the complexity of our environment, it is difficult to obtain the transition probability, thus making model-based methods inapplicable to our problem. On the other hand, the model-free methods rely on past experience without making any assumptions about the environment. Therefore, the model-free RL methods are suitable for handling the above-formulated problem in our case.

\section{Multi-Agent Soft Actor-Critic with Graph Convolutional Networks} \label{masac}
Soft Actor-Critic differs from other RL methods in that it aims not only to maximize discounted cumulative reward but also to maximize entropy, which is used to measure the randomness in the policy \cite{haarnoja2018soft}. In this section, we propose SAC-GCN, a multi-agent version of SAC with graph convolutional networks and self-attention mechanism, to deal with the resource allocation problem in the metaverse. 
\subsection{Multi-Agent Soft Actor-Critic}
The objective of SAC-GCN is to find the optimal joint policy $\pi^*$ to maximize the expected discounted return and its entropy simultaneously
\begin{equation}
\pi^* = \argmax_{\pi}\ \mathbb {E}_\pi \bigg [ \sum^{\infty}_{k=0} \gamma^k \bigg (r_{t+k+1} + \alpha H(\pi(\cdot|\boldsymbol{s}_{t+k})) \bigg )\bigg ]
\end{equation}
where the entropy term $H(\pi(\cdot|\boldsymbol{s}_t))$ is calculated by $H(\pi(\cdot|\boldsymbol{s}_t)) = \mathbb {E}_{a}[-log(\pi(\boldsymbol{a}|\boldsymbol{s}_t))]$ and the temperature parameter $\alpha$ determines the importance of the entropy value compared to the reward. The soft state value function is denoted as 
\begin{equation}
\begin{split}
Q( & \boldsymbol{s}, \boldsymbol{a})=\\ & \mathbb {E}_{\pi} \bigg [ V^\pi(\boldsymbol{s}) + \alpha \sum^{\infty}_{k=1} \gamma^k H(\pi(\cdot|\boldsymbol{s}_{t+k}))|\boldsymbol{s}_t = \boldsymbol{s}, \boldsymbol{a}_t = \boldsymbol{a} \bigg ]
\label{soft state action value}
\end{split}
\end{equation}
where $V^\pi(\boldsymbol{s})$ is the expected future rewards defined in Eq. \ref{state value}. The soft state value function is then given based on the soft Q-function \cite{haarnoja2018soft}
\begin{equation}
V(\boldsymbol{s}) = \mathbb {E}_{\boldsymbol{a} \sim \pi(\cdot|\boldsymbol{s})}[Q(\boldsymbol{s}, \boldsymbol{a}) - \alpha \log(\pi(\boldsymbol{a}|\boldsymbol{s})) |\boldsymbol{s}_t = \boldsymbol{s}, \boldsymbol{a}_t = \boldsymbol{a}]
\label{soft state value}
\end{equation}

The multiple agents of SAC-GCN are organized by the centralized training and decentralized execution framework, in which agents are trained in a centralized offline manner and execute online with decentralized information. Specifically, during the training process, the centralized critic of each agent in SAC-GCN can obtain global states and joint actions from all the agents. The policy of the agent is then updated by the centrally learned value function. Once the training is finished, decisions are made solely on the actor's observations. For each agent $i$ in SAC-GCN, we maintain two Q-networks (parameterized by $\phi^i_1$ and $\phi^i_2$) to avoid overestimation of Q-function values and two target networks (parameterized by $\overline {\phi}^i_1$ and $\overline {\phi}^i_2$) to stabilize the training process \cite{van2016deep}. The target networks are updated in a soft manner
\begin{equation}
\overline {\phi}^i_j \gets \tau \phi^i_j + (1 - \tau) \overline {\phi}^i_j, j = 1, 2
\label{update target}
\end{equation}
where $\tau$ is the step size. The policy network of agent $i$ is parameterized by $\theta^i$ and all the policy parameters are packed together as the joint policy denoted as $\boldsymbol{\theta} = (\theta^1, ..., \theta^N)$. Due to the independence of each $\theta^i$, the distribution of the joint action under $\boldsymbol{\theta}$ is calculated by
\begin{equation}
\begin{split}
\pi_{\boldsymbol{\theta}}(a_{t}|s_{t}) = \prod_{i=1}^{n} \pi_{\theta^i}(a^i_{t}|o^i_{t})
\end{split}
\end{equation}

To obtain the optimal trade-off between the expected discounted return and the expected entropy at each state, we employ the soft state value (Eq.\ref{soft state value}) as the target of our policy $\pi_{\theta^i}$
\begin{equation}
\begin{split}
\max_{\theta^i}\ \mathbb {E}_{\boldsymbol{a}_t \sim \pi_{\boldsymbol{\theta}}}[Q_{\phi}(\boldsymbol{s}_t, \boldsymbol{a}_t) - \alpha \log(\pi_{\theta^i}(a^i_t|o^i_t))]
\end{split}
\end{equation}
where the expected value is approximated by sampling. However, direct sampling from the distribution $\pi_{\theta^i}(\cdot|o^i_t)$ which is parameterized by the target function is not differentiable. Thus, the target function is unable to be updated by backpropagation in neural networks. We follow the reparameterization trick in \cite{haarnoja2018soft} and obtain the samples by a squashed Gaussian policy
\begin{equation}
\tilde{a}_{\theta^i}(o^i_t, \xi^i_t) = \tanh(\mu_{\theta^i}(o^i_t) + \sigma_{\theta^i}(o^i_t) \odot \xi^i_t), \xi^i_t \sim \mathcal{N}(0, 1)
\label{gaussain}
\end{equation}
where $\mu_{\theta^i}(o^i_t)$ and $\sigma_{\theta^i}(o^i_t)$ are the mean and standard deviation of a Gaussian distribution generated by the policy network. The target of the policy network is then given by samples from the replay buffer $\mathcal{B}$
\begin{equation}
\begin{split}
J_\pi(\theta^i) &= \mathbb {E}_{\boldsymbol{s}_t \sim \mathcal{B}, \xi_i \sim \mathcal{N}}[\min_{j=1,2}Q_{\phi_j}(\boldsymbol{s}_t, \tilde{\boldsymbol{a}_t}) - \alpha \log(\pi_{\theta^i}(\tilde{a}^i_t|o^i_t))]
\end{split}
\label{target policy}
\end{equation}
where $\tilde{a}^i_t$ represents $\tilde{a}_{\theta^i}(o^i_t, \xi^i_t)$ in Eq. \ref{gaussain} and $\tilde{\boldsymbol{a}}_t$ is the joint sampled action by a Gaussian policy, denoted as $\tilde{\boldsymbol{a}}_t = (\tilde{a}^1_t, \tilde{a}^2_t, ..., \tilde{a}^N_t)$.

\begin{figure*}[htbp]
\centering
\includegraphics[width=0.95\textwidth]{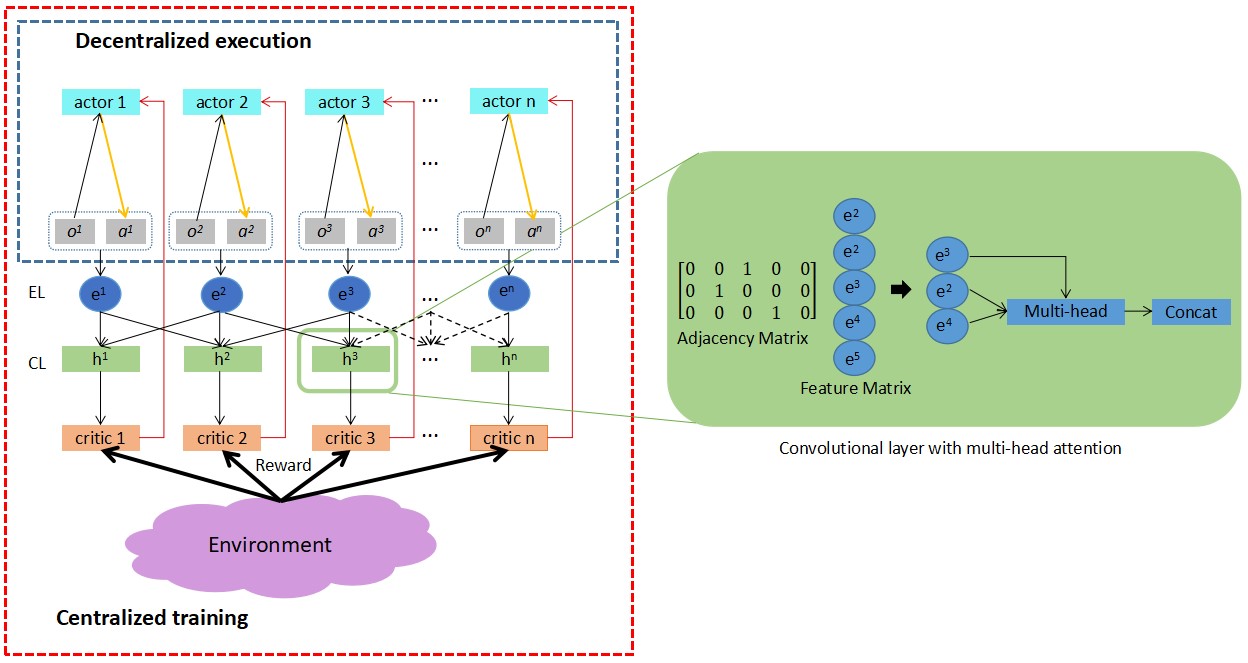}
\caption{The structure of the neural networks in SAC-GCN, where ``EL" and ``CL" represent the encoder layer and the convolutional layer, respectively. Each agent has a centralized critic dealing with joint states and actions during training and a decentralized actor taking actions based on local observations in the execution process. The convolutional graph networks with multi-head attention are employed to obtain the critic's final inputs.} 
\label{network structure}
\end{figure*}

In the Q-network, the loss function is to minimize the temporal difference error, defined as
\begin{equation}
\begin{split}
J_Q(\phi^i_{1,2}) = \mathbb {E}_{(\boldsymbol{s}_t, \boldsymbol{s}_{t+1}) \sim \mathcal{B}, (\xi^i_t, \xi^i_{t+1}) \sim \mathcal{N}}[(Q_{\phi^i_{1,2}}(\boldsymbol{s}_t, \tilde{\boldsymbol{a}}_t) - y')^2] 
\end{split}
\label{target value}
\end{equation}
where the target is given by
\begin{equation}
\begin{split}
y' = r_t + \gamma (\min_{j=1,2}Q_{\overline \phi_j}(\boldsymbol{s}_{t+1}, \tilde{\boldsymbol{a}}_{t+1})- \alpha \log(\pi_{\theta^i}(\tilde{a}^i_{t+1}|\boldsymbol{s}_{t+1})))
\end{split}
\end{equation}
Here $\min_{j=1,2}Q_{\overline \phi_j}(\boldsymbol{s}_{t+1}, \boldsymbol{a}_{t+1})$ denotes the minimum of the two target Q-networks. 
With the defined target functions for the value network (Eq. \ref{target value}) and the policy network (Eq. \ref{target policy}), the parameters  are updated by gradient descent and gradient ascent, respectively. 

\subsection{Graph Convolutional Networks}
During the above training process of SAC-GCN, the centralized critics are fed with global states from all agents making it hard to get valuable information. To tackle this problem, we employ graph convolutional networks to explore the hidden graph of agents. More specifically, the multi-agent metaverse environment is modeled as a graph where each agent can be represented as a node. The nodes' features are derived from local observations which are the conditions of communication resources and computation resources at the edge. In the metaverse, one user's movement/interaction would likely impact its neighboring users, whose resource requests would accordingly change. Therefore, we add edges between a node and its neighboring nodes in the graph to represent their mutual influence. 

Each agent of SAC-GCN consists of three components: an encoder layer, a convolutional layer, and SAC networks, as shown in Fig. \ref{network structure}. The encoder layer encodes the local observation and action to obtain a high-dimensional representation which is further used as the feature of a node. The convolutional layer takes the features from all the nodes within the same neighborhood as input and outputs a latent representative feature vector to represent these nodes. We construct a feature matrix $F$ where the feature vectors of all the nodes are organized row by row to represent the multi-agent environment. An adjacency matrix $A$ is also built where the first row is the one-hot representation of the index of node $i$ followed by the same representation of the indexes of neighboring nodes. Take agent 3 as an example (shown in Fig. \ref{network structure}), the adjacency matrix starts with the one-hot representation of number 3 ($[0,0,1,0,0]$) followed by that of neighboring agent 2 and agent 4. The final input of the critic $i$ is then calculated by $A \times F$ with the self-attention mechanism discussed in the next subsection.

\begin{algorithm} 
    \renewcommand{\algorithmicrequire}{\textbf{Input:}}
	\renewcommand{\algorithmicensure}{\textbf{Output:}}
	\caption{SAC-GCN} 
	\label{alg1} 
	\begin{algorithmic}[1]
		\STATE Initialize two soft Q-networks with parameters $\phi^i_1$, $\phi^i_2$ and a policy network with parameters $\theta^i$, for each agent $i$
		\STATE Initialize two target networks with $\overline {\phi^i_1} \gets \phi^i_1$ and $\overline {\phi^i_2} \gets \phi^i_2$, for each agent $i$
		\FOR{episode $e = 1, 2, ...$}
		    \FOR{time step $t = 1, 2, ..., T$}
		        \STATE Obtain the original observation $o^i_t$ for each agent and current global state $\boldsymbol{s}_t$
                \FOR{each agent $i$}
                    \STATE Take an action $\tilde{a}^i_t$ generated by Eq. \ref{gaussain} 
                    \STATE Receive shared reward $r_t$ and observe $o^i_{t+1}$
                \ENDFOR
                \STATE Obtain joint action $\tilde{\boldsymbol{a}}_t$, and next global state $\boldsymbol{s}_{t+1}$ 
                \STATE Store the tuple sample ($\boldsymbol{s}_t, \tilde{\boldsymbol{a}}_t, r_t, \boldsymbol{s}_{t+1}$) into the replay buffer $\mathcal{B}$
                \FOR{each agent $i$}
                    \STATE Get a random batch of samples from $\mathcal{B}$
                    \STATE Apply attention mechanism (Eq. \ref{attention}) to get the inputs of each critic $h_t$ and $h_{t+1}$ based on global states $\boldsymbol{s}_t$ and $\boldsymbol{s}_{t+1}$
                    \STATE Update Q-networks by minimizing the loss function (Eq. \ref{target value}) with $\boldsymbol{s}_t$ replaced by $h_t$ and $\boldsymbol{s}_{t+1}$ replaced by $h_{t+1}$
                    \STATE Update policy networks by maximizing the target function (Eq. \ref{target policy}) with $\boldsymbol{s}_t$ replaced by $h_t$
                    \STATE Update target Q-networks by Eq. \ref{update target}
                    
                \ENDFOR
            \ENDFOR
	    \ENDFOR
	\end{algorithmic} 
\end{algorithm}
 
\subsection{Self-Attention Mechanism}
The self-attention mechanism is employed within the convolutional layer for agents to give different attention to neighboring features. The intuition behind this is that in the multi-user metaverse environment, the impacts of actions taken by other agents are different for a specific agent. For example, when deciding how to adjust the target bit rate, an agent should pay more attention to the agents with more network resources. To apply self-attention mechanism, we first transform the feature of node $i$ to a ``Query" calculated by $Q = W_qe^i$, where $e^i$ is the encoded feature of node $i$. Each neighboring node $j$ is then transformed to a ``Key$^j$" and a ``Value$^j$" by $K^j = W_ke^j$ and $V^j = W_ve^j$, respectively. The similarity between the Query and Key$^j$ is calculated using the cosine similarity denoted as
\begin{equation}
S(Q, K^j) = \frac{Q\cdot K^j}{||Q||\cdot ||K^j||} 
\end{equation}
The attention weight for Key$^j$ is obtained by the softmax function
\begin{equation}
\alpha^j = \frac{exp(S(Q, K^j))}{\sum_j exp(S(Q, K^j))} 
\end{equation}
Finally, the neighboring feature for a node is calculated by a weighted sum of the values using the above attention weights
\begin{equation}
Attention(Q, K, V) = \sum_j \alpha^j V^j
\label{attention}
\end{equation}
Moreover, to obtain better performance, multiple attention heads \cite{vaswani2017attention} with multiple sets of parameters $(W_k, W_q, W_v)$ are employed to perform attention allocation multiple times in parallel. We then concatenate outputs from all heads to a single vector as the final representation of the neighboring feature for node $i$. 

A detailed description of the proposed SAC-GCN can be found in Algorithm \ref{alg1}. At each iteration, we first randomly obtain a batch from the replay buffer $\mathcal{B}$ for each agent. We then apply GCN with multi-head attention to calculate the inputs of the centralized critic based on Eq. \ref{attention}. The parameters of the Q-networks, the policy networks, and the target Q-networks are updated in steps 15-17, respectively. Since we employ the offline centralized training and online decentralized execution mechanism, there is no extra delay during execution once the training process is finished. Therefore, the well-trained SAC-GCN model can be used for the delay-sensitive metaverse.


\section{Case Study: Resource Allocation in A Virtual City Park Metaverse} \label{case}
In this section, we describe a case study of how to adaptively allocate communication and computation resources for users in a virtual city park metaverse. In more detail, we describe the architecture of our smart resource allocation system, followed by the three evaluation baselines that we use. The offline training results of these deep reinforcement learning (DRL) methods are demonstrated in the end.  
\begin{figure*}[htbp]
\centering
\includegraphics[width=0.95\textwidth]{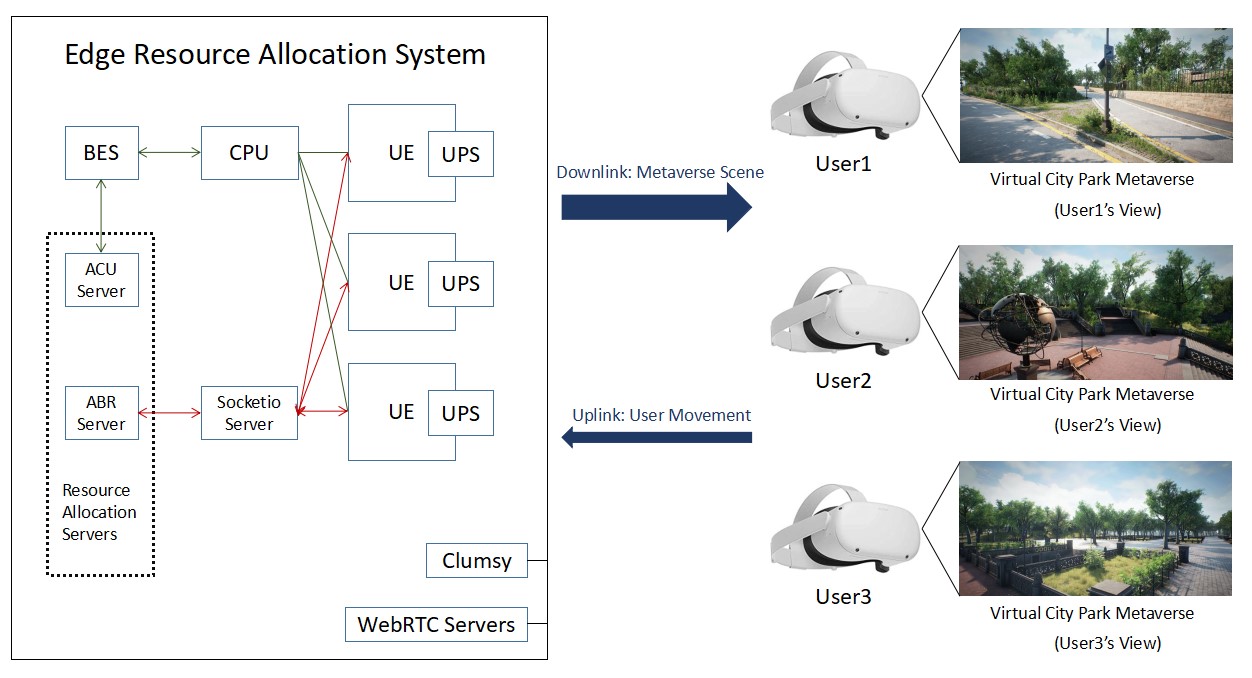}
\caption{The architecture of our proposed resource allocation system for the metaverse. Two resource allocation servers, ABR and ACU, are responsible to control the users' consumption of network resources and CPU resources, respectively. Moreover, with this system, different scenes of the virtual city park are displayed to users wearing VR headsets.} 
\label{system architecture}
\end{figure*}
\subsection{System Setup}
The proposed resource allocation system is built on a desktop computer with the Windows 10 operating system, an Intel Core i9-11900F processor as its central processing unit (CPU), 64 GB of random-access memory (RAM), and an NVIDIA GeForce RTX 3090 graphics processor. The virtual city park\footnote{https://www.unrealengine.com/marketplace/en-US/product/city-park-environment-collection} in the metaverse is designed with Unreal Engine (UE). As shown in Fig. \ref{system architecture}, the desktop computer connects to a TP-Link TL-WR1043ND Wi-Fi router by an Ethernet cable. We run multiple instances of UE simultaneously to provide users with different scenes. With the Unreal Pixel Streaming (UPS) plugin of Unreal Engine, the frames rendered by the GPU-enabled desktop can be transmitted to remote users with VR headsets (Meta Quest 2 in our case) through WebRTC peer-to-peer communication protocol \cite{nurminen2013p2p}. 

Here is how the resource allocation system works. UPS first finds the IP addresses of the users' VR headsets through WebRTC servers, and then creates a wireless communication session between UPS and VR headsets. The communication session allows users' inputs, such as movement and commands, to be transmitted to the UE instances as mice flows through the uplink traffic, while rendered frames are transmitted back to the users as elephant flows through the downlink traffic. The instances of UE and the adaptive bit rate (ABR) server are connected by Socket.io\footnote{https://socket.io/} which enables real-time bidirectional event-based communication. Therefore, the ABR server receives the network information from UE and determines the bit rate selection for each user. Regarding computation resource allocation, the adaptive CPU usage (ACU) server detects the CPU performance and employs BES\footnote{https://mion.yosei.fi/BES/} to throttle the CPU usage for the instances of UE individually. In addition, Clumsy\footnote{ https://github.com/jagt/clumsy}, a Windows network controller, is used to simulate the dynamics of the network by controlling delay, bandwidth, and random packet loss rate.

In our experiments, the default settings of the virtual city park are demonstrated as follows: bandwidth = 300Mbps, network delay = 20ms, packet loss rate = 0.5\%, resolution = 2048x1080, frame rate = 60fps, and available CPU percentage = 80\%. Moreover, we explored the impact of one specific parameter with a wide range of values. In more detail, we set delay between 10ms to 200ms, packet loss rate between 0.5\% to 8\%, bandwidth between 50Mbps to 400Mbps, and available CPU between 10\% to 90\%. 

\subsection{Evaluation Baselines} \label{baselines}
To demonstrate the effectiveness of the proposed SAC-GCN in resource allocation for the metaverse, we compared it with the following methods:
\begin{itemize}
\item DQN: It is one of the representative DRL methods which combines the advantages of Q learning and neural networks \cite{mnih2013playing}. Specifically, it employs neural networks to replace the Q table to make it able to deal with a large number of states. Previous research has demonstrated the effectiveness of DQN for solving the problem of resource allocation at the edge \cite{wang2019smart, wu2021hybrid}. 
\item Independent SAC (ISAC): This method is a totally decentralized form of multi-agent SAC where each agent is an independent learner taking the rest of the agents as part of the environment. Despite its various theoretical shortcomings, ISAC is appealing compared to MASAC as each agent only requires its local observations without the communication and scalability problem.
\item GCC with Greedy (GCC-G): It employs Google Congestion Control (GCC) \cite{carlucci2016analysis} to control the bit rate while allowing the instances of UE to greedily utilize the computation resources without any limitation. This is similar to the way that real-time communication tools powered by WebRTC work.
\item BBR with Greedy (BBR-G): It is also a congestion control algorithm developed at Google. BBR determines the maximum bandwidth by sending more data than the capacity of the network, and when the delay increases after increasing the volume of data sent, the maximum bandwidth is determined. The minimum delay is calculated by sending data below the network capacity. The RTT obtained is the minimum delay when the delay does not drop after decreasing the sending volume. Similar to GCC-G, it also allows the instances of UE to greedily use the computation resources.
\end{itemize}
\subsection{Offline Training}
As mentioned previously, for each agent in SAC-GCN (Fig \ref{network structure}), the neural network consists of two encoding layers and a convolutional layer apart from the actor network and the critic network. Both the soft-Q networks and the policy network have an input layer, two hidden layers activated by the rectified linear unit (ReLU) function, and an output layer. All the key parameters we set up during the implementation of SAC-GCN are summarized in Table \ref{paras}. To ensure fairness, all the methods are configured with the same parameters.

\begin{table}[htbp]
\centering
\caption{Parameter Specifications in the Experiments}
\label{paras}
\begin{adjustbox}{width=0.46\textwidth}
\begin{tabular} {|c|c|}
\hline
\textbf{Parameters} & \textbf{Value}\\
\hline
Optimizer & Adam\\
\hline
Number of neurons in hidden layer & 128\\
\hline
Reward discount factor & 0.99 \\
\hline
Replay buffer size & 5000\\
\hline
Learning rate for updating Q-networks & 0.0001\\
\hline
Learning rate for updating policy network & 0.0005\\
\hline
Batch size & 64 \\
\hline
Coefficient for updating target Q-networks & 0.01\\
\hline
Entropy temperature parameter & 0.2\\
\hline
Metaverse scene satisfaction level factor & 1\\
\hline
Choppiness penalty factor & 0.2\\
\hline
Latency penalty factor & 0.05\\
\hline
Instability penalty factor & 0.5\\
\hline
Overall QoE factor & 2\\
\hline
Communication resource balancing factor & -0.6\\
\hline
Computation resource balancing factor & -0.6\\
\hline
Number of users in the metaverse & 3\\
\hline
\end{tabular}
\end{adjustbox} 
\end{table}

\begin{figure}[htbp]
\centering
\includegraphics[width=0.51\textwidth]{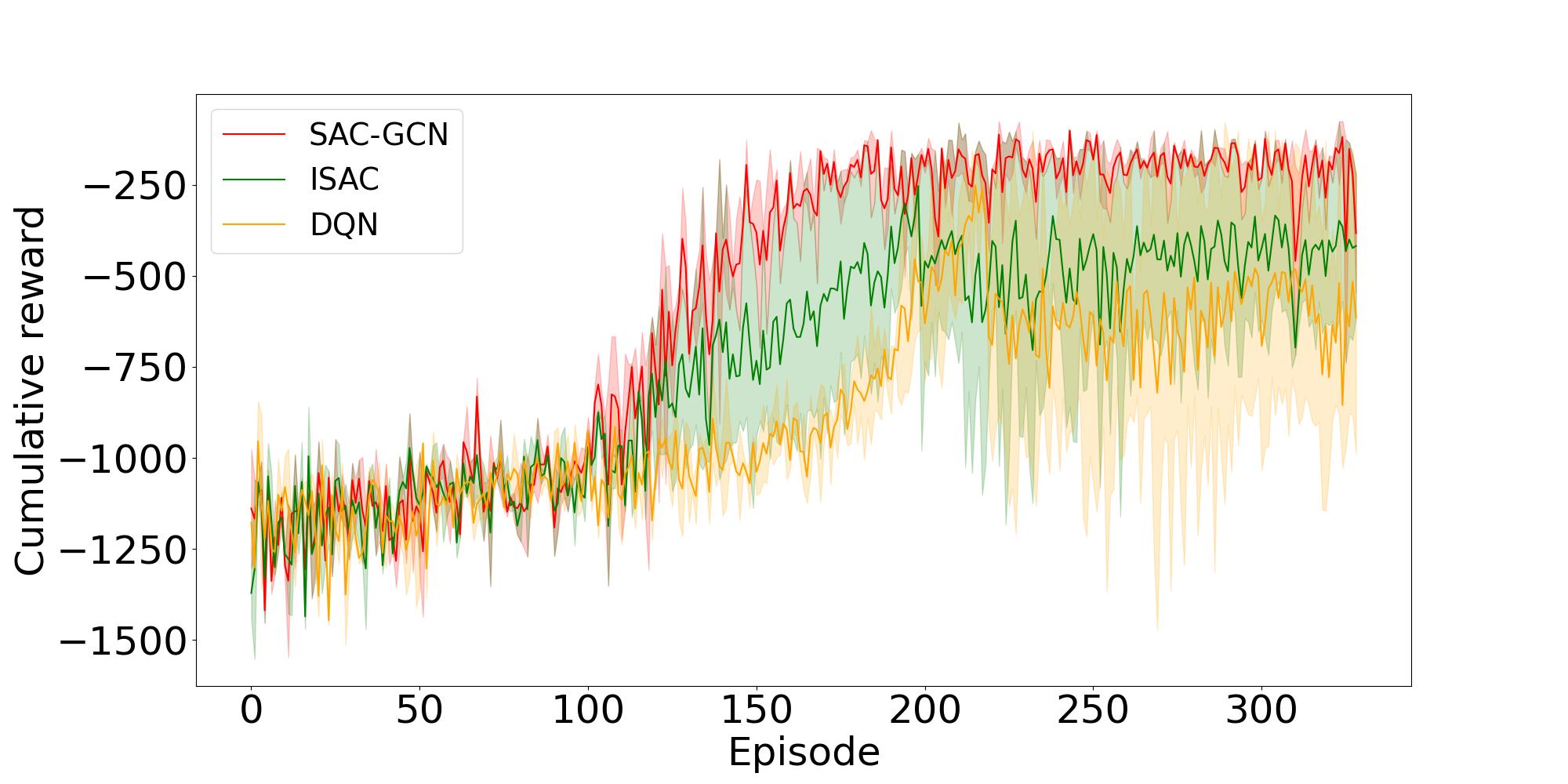}
\caption{Cumulative reward versus episode (40s for an episode) for different DRL methods during the offline training process.} 
\label{training}
\end{figure}

Unlike other typical RL problems, there is no clear definition of an ``episode" in the metaverse. We empirically set a period of time (i.e., 40s) as an episode during which agents take actions every second. We train each method under different network conditions and rendering settings in order to adapt to the dynamic metaverse environment. The learning curves of the three DRL methods in terms of cumulative reward are demonstrated in Fig. \ref{training}. We notice that at the beginning of the training, the cumulative rewards of three methods are all low without obvious difference between them. This is due to the fact that these DRL methods cannot learn a stable and efficient policy for resource allocation with a small amount of data by interacting with the environment. As the number of training episode increases, the cumulative rewards of SAC-GCN and ISAC are higher than that of DQN. Our explanation is that they take a greater variety of actions during exploration with a higher entropy of the strategy, which accelerates the learning process and minimizes the risk of a local optimum. Moreover, the globally learned strategy based on the centralized training and decentralized execution structure improves the performance of SAC-GCN compared to ISAC. Therefore, SAC-GCN reaches the highest cumulative reward at the end. 

\begin{figure*}[htbp] 
\centering
\subfigure[]{
\includegraphics[width=0.48\textwidth, height=0.25\textwidth]{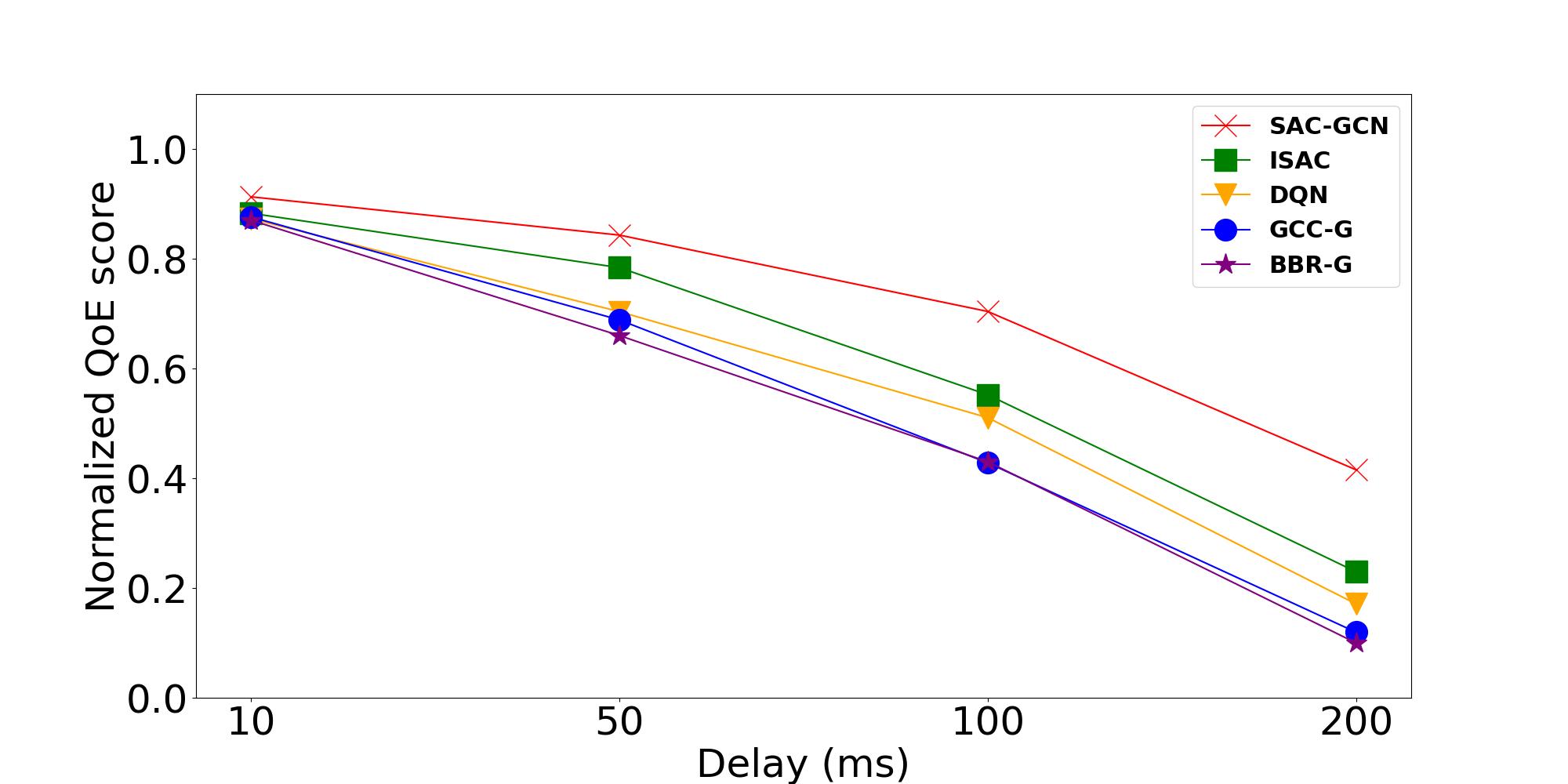}}
\subfigure[]{
\includegraphics[width=0.48\textwidth, height=0.25\textwidth]{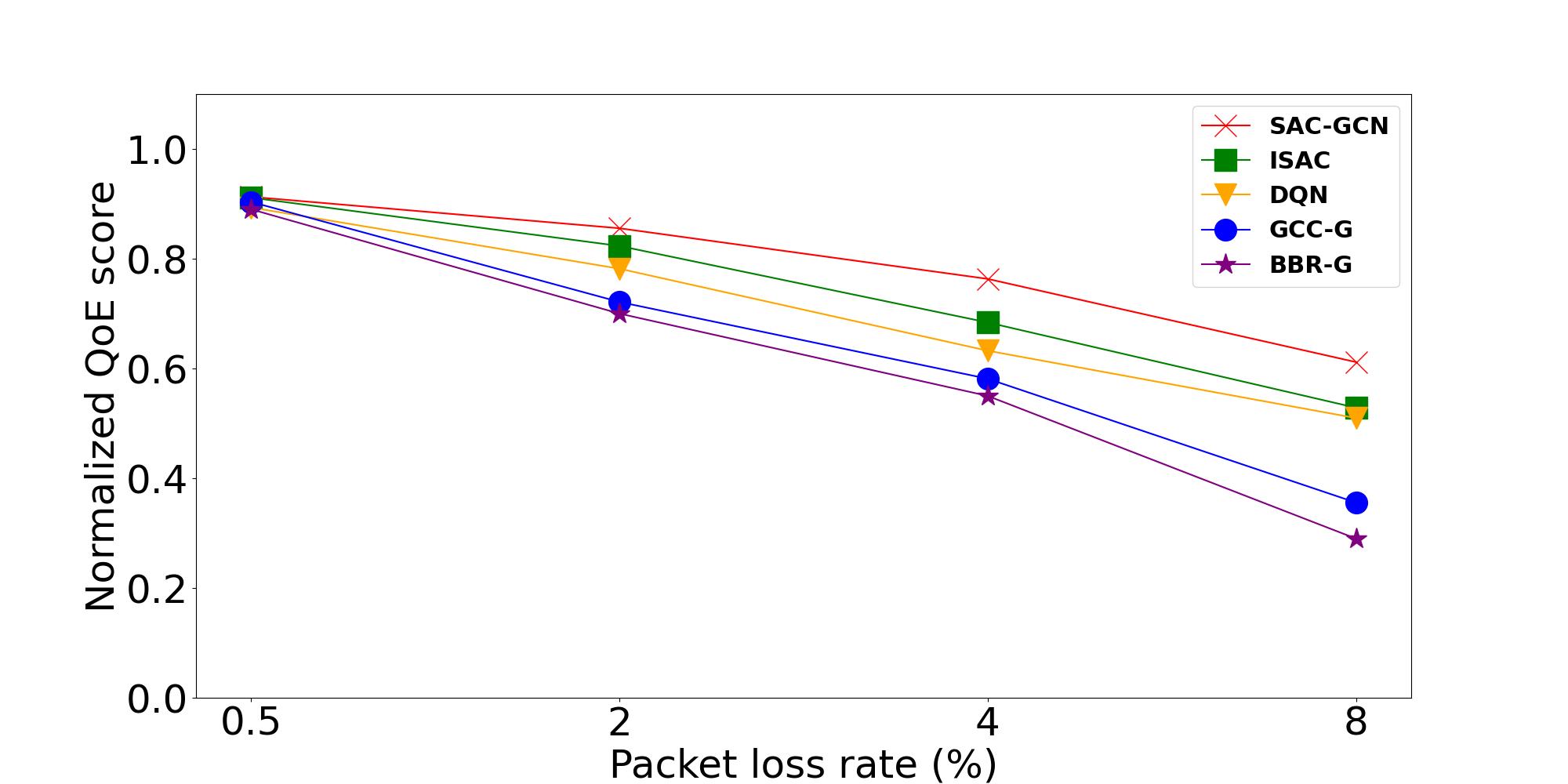}}
\subfigure[]{
\includegraphics[width=0.48\textwidth, height=0.25\textwidth]{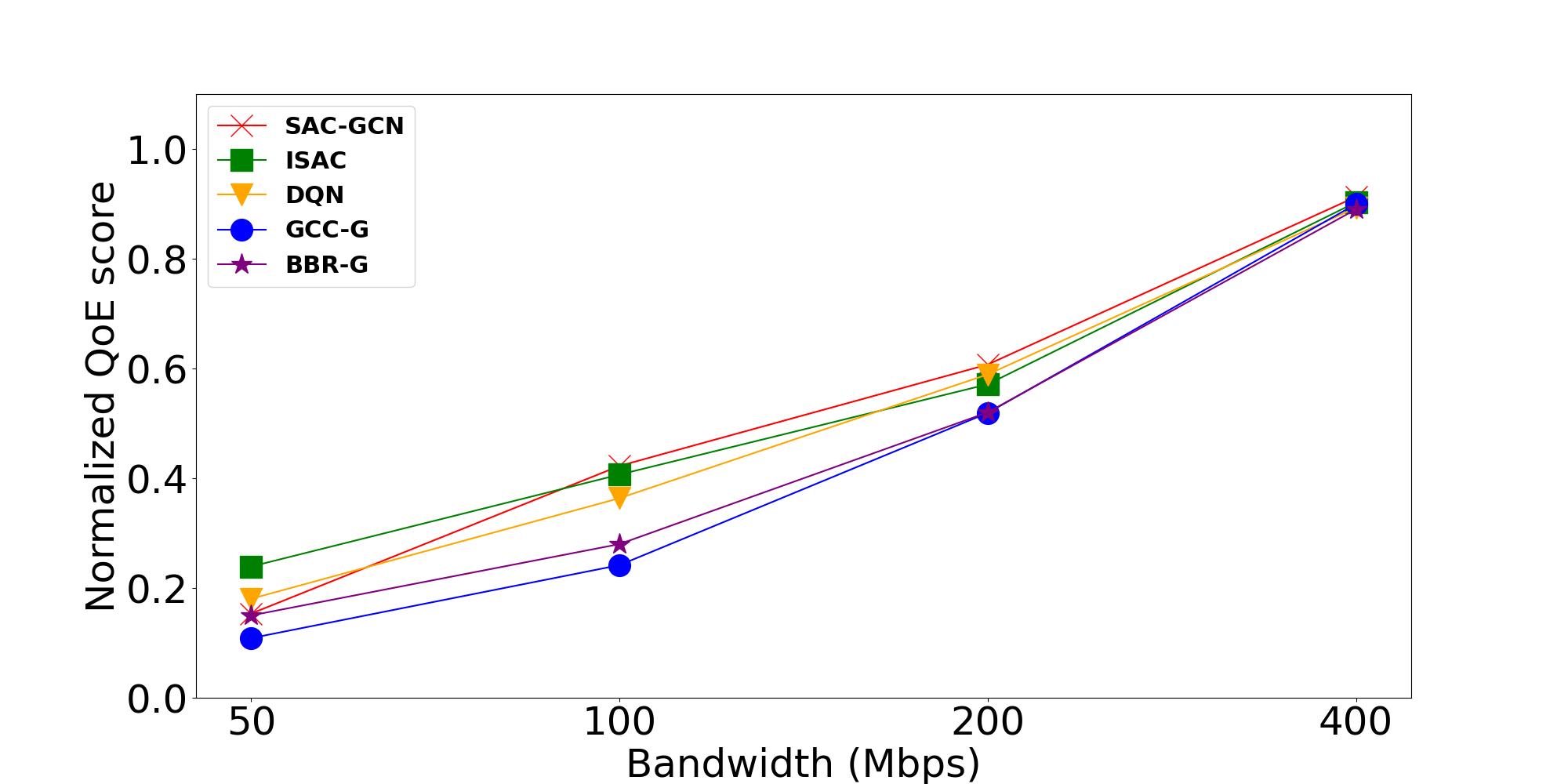}}
\subfigure[]{
\includegraphics[width=0.48\textwidth, height=0.25\textwidth]{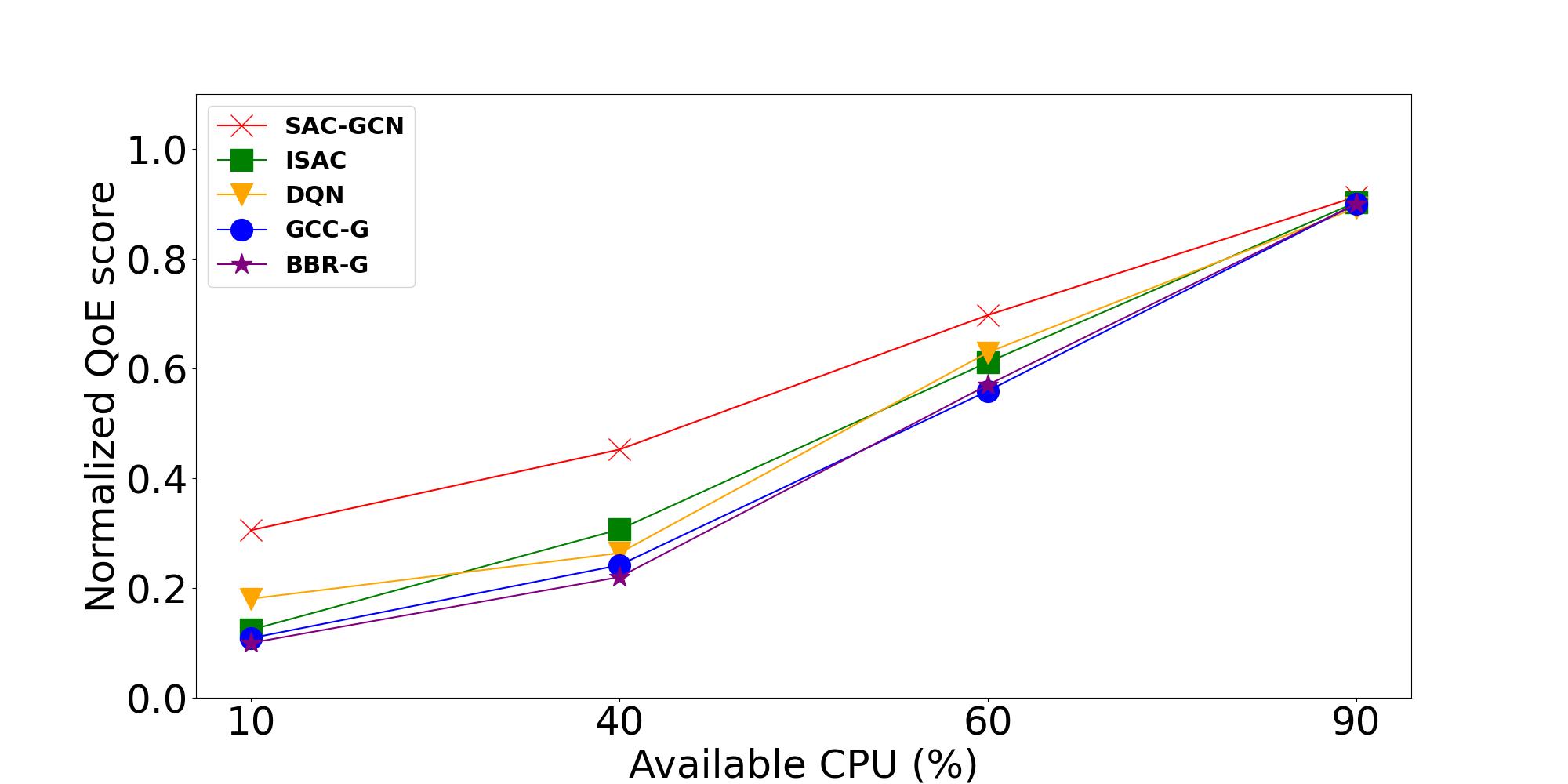}}
\caption{Overall QoE score comparison of the five resource allocation methods under different settings: (a) delay (10ms, 50ms, 100ms, and 200ms), (b) packet loss rate (0.5\%, 2\%, 4\%, and 8\%), (c) bandwidth (50Mbps, 100Mbps, 200Mbps, and 400Mbps), and (d) available CPU (10\%, 40\%, 60\%, and 90\%).}
\label{qoe_comp}
\end{figure*}

\section{Online Experimental Results and Analysis} \label{analysis} 
We compared the performance of our proposed SAC-GCN with the four evaluation baselines for resource allocation (Subsection \ref{baselines}): DQN, ISAC, GCC-G, and BBR-G. The performance metrics in our experiments are: i) overall user QoE : the sum of QoE score for each user in the metaverse, where the definition of QoE is given in the reward function of SAC-GCN (Subsection \ref{user qoe}); ii) balance of resource allocation: balance of communication resource is represented by variance of users' bit rates and balance of computation resource is denoted by variance of users' CPU usage percentages; iii) resource utilization rate: communication resource utilization rate is calculated by the sum of users' bit rates divided by the assigned total bandwidth and computation resource utilization rate is determined by the sum of users' CPU usage percentages divided by the total available CPU. Note that the values of overall QoE, variance of resource allocation, and resource utilization rates are all normalized between 0 to 1, respectively. 

\subsection{Overall User Quality of Experience Analysis}
We first evaluated the performance of all resource allocation methods regarding overall QoE under various network delays (shown in Fig. \ref{qoe_comp} (a)). It is noted that the network delay is the extra time we add before the packets are transmitted which does not include the time spent on the network transmission itself. The overall QoE scores of all methods show a similar downward trend as the network delay increases. Furthermore, the QoE scores decrease more with the increase of the network delay. It is because the Motion-to-Photon (MTP) latency, which describes the time between the movement of a tracked object and its corresponding image displayed on the screen, gets higher with the increasing extra network delays. The high MTP latency may cause a significant loss of performance with cybersickness in the interactive metaverse. By employing graph neural networks with self-attention mechanism, agents of SAC-GCN pay more attention to neighboring agents with high delays when making decisions. Thus, SAC-GCN can always choose appropriate bit rates for each user to alleviate the impact of high delays on QoE. We observe that SAC-GCN outperforms the other methods under different network delays especially when the delay gets large. For example, when the network delay is 100ms, SAC-GCN reaches 27\%, 37\%, 63\%, and 64\% improvements over ISAC, DQN, GCC-G, and BBR-G, respectively.


We then examined the impact of network packet loss rate on QoE, as illustrated in Fig. \ref{qoe_comp} (b). Note that the network packet loss rate is manually set by randomly dropping some packets before they are transmitted and it coexists with the packet loss caused by network congestion. We observe that the increased packet loss rate degrades user experience for all the methods and it gets worse when the packet loss rate is high. It is because when the packet loss rate is low, it only causes a small number of frames dropped which is not easy to be noticed. However, as the packet loss rate increases, user experience is largely impacted by the choppiness caused by lost frames. The increased packet loss rate has the biggest influence on BBR-G. More specifically, the QoE of BBR-G decreases 0.19 when packet loss rate increases from 0.5\% to 2\%, 0.15 (from 2\% to 4\%), and 0.26 (from 4\% to 8\%), which are much larger than the other three DRL methods. SAC-GCN always has the best performance when dealing with different packet loss rates. The reason is that the employed graph convolutional networks can help agents in SAC-GCN notice the dramatic changes of packet loss rate for neighboring agents. They can take quick actions based on the underlying interplay between agents. Therefore, agents of SAC-GCN are able to cooperatively reduce the impact of packet loss rate, thereby reaching the highest QoE.


The impact of total available network bandwidth on overall QoE was studied in Fig. \ref{qoe_comp} (c). It is observed that when the bandwidth is extremely low (50Mbps and 100Mbps), all methods end up with low QoE scores. The reason is that some of the packets for the metaverse scenes are either delayed or dropped in such conditions. Therefore, with insufficient bandwidth, metaverse users may all suffer from cybersickness caused by high MTP and choppiness due to a large number of dropped frames. We also perceive that when the bandwidth is insufficient (50Mbps), SAC-GCN performs worse than ISAC. Note that ISAC simply assumes that all agents are independent which makes it lack convergence guarantees and not theoretically sound in the multi-agent setting. However, agents of ISAC make decisions only by local information during the training and inference. Thus, they do not need to wait for information from other agents relating to communication and coordination costs (even severe when the bandwidth resource is very insufficient) which provides a quicker response compared with SAC-GCN. We think this is the reason why ISAC outperforms SAC-GCN in this case. As the bandwidth grows, the performance of all methods increases to varying degrees. When the total bandwidth is sufficient for all users (400Mbps), all methods achieve high QoE.


\subsection{Resource Allocation Balancing Analysis}
\begin{figure}[htbp] 
\centering
\subfigure[]{
\includegraphics[width=0.48\textwidth, height=0.25\textwidth]{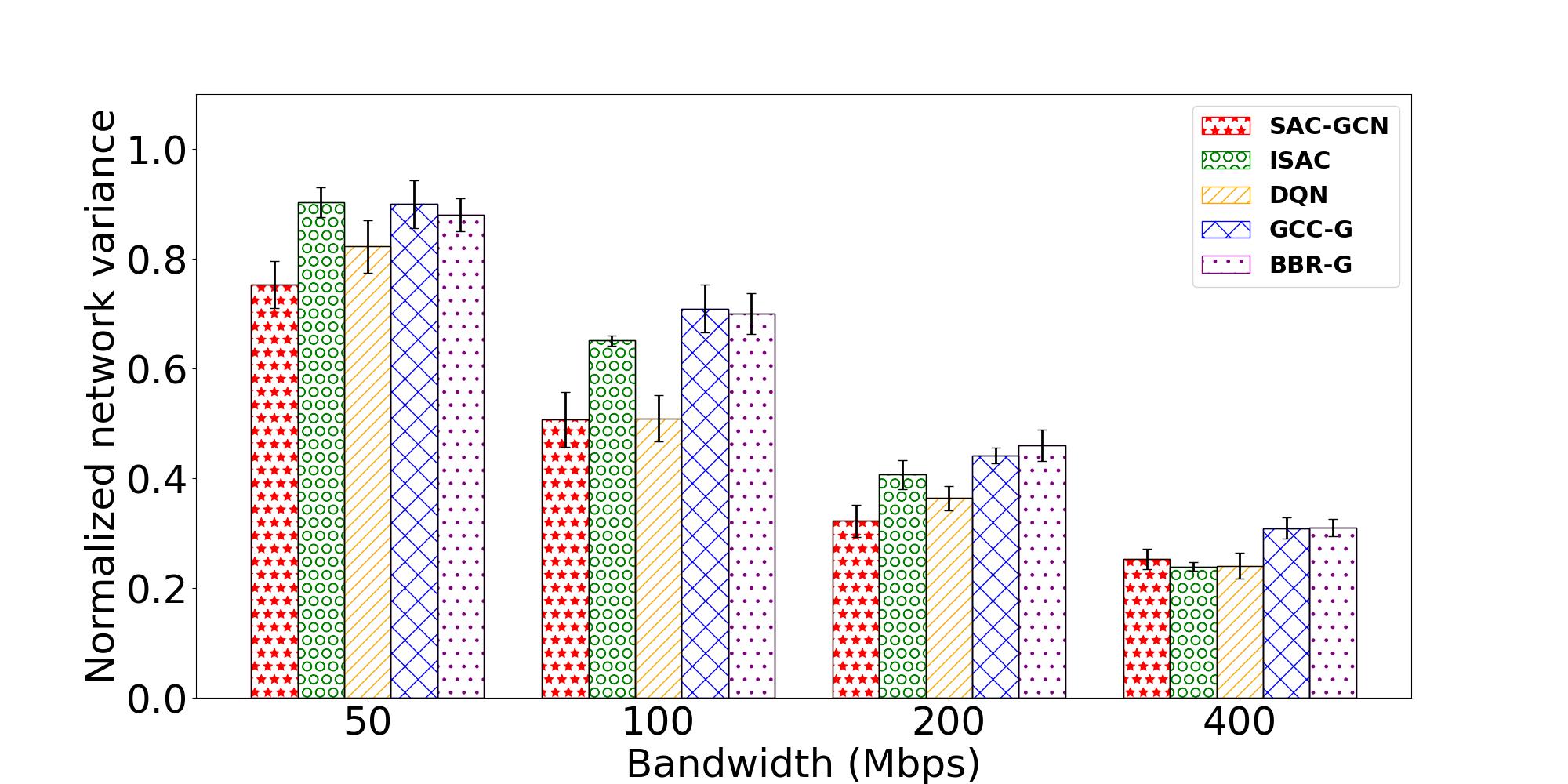}}
\subfigure[]{
\includegraphics[width=0.48\textwidth, height=0.25\textwidth]{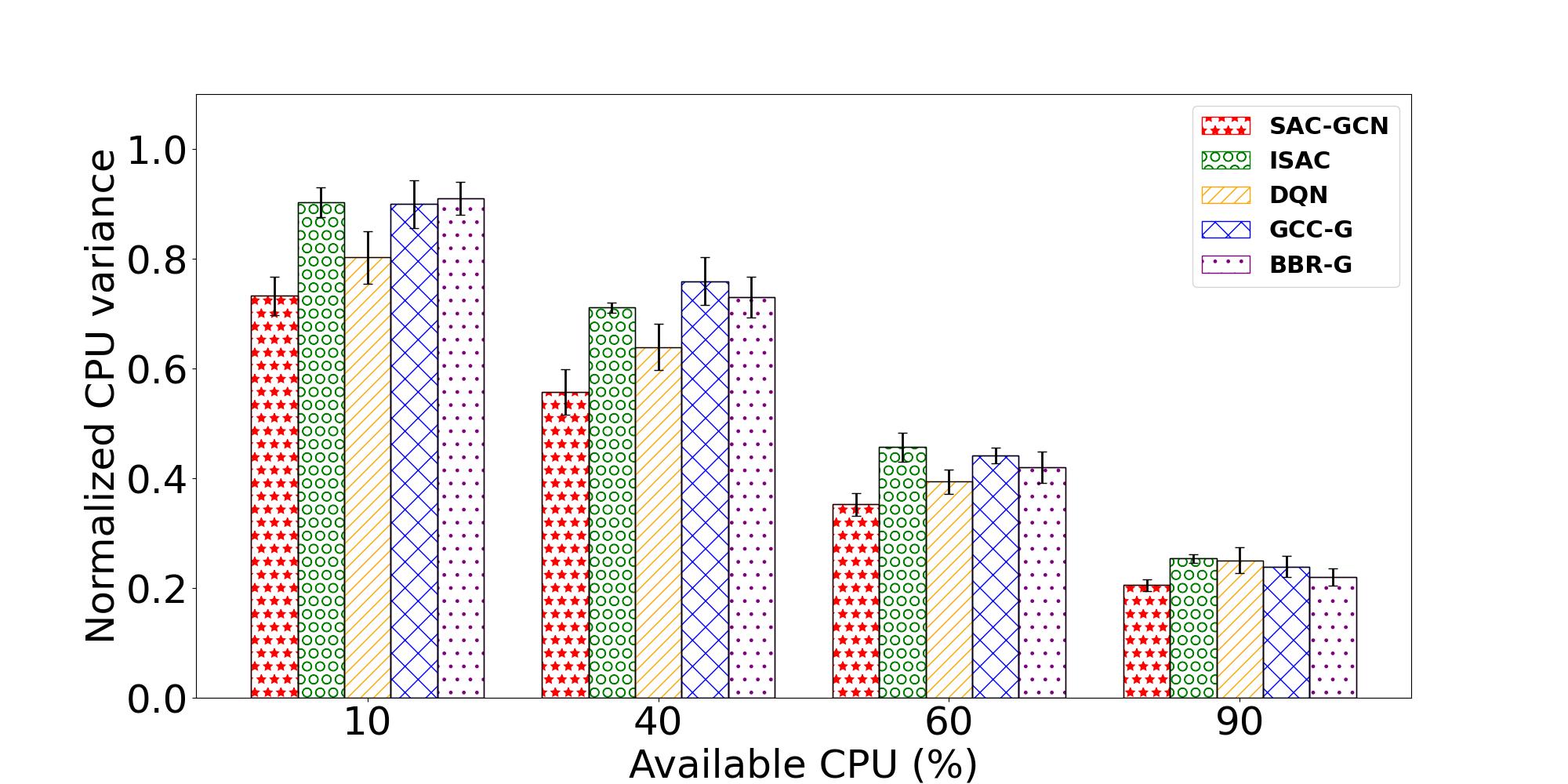}}
\caption{Comparison of resource allocation balancing under different settings: (a) impact of network bandwidth on balance of network and (b) impact of available CPU percentage on balance of computation allocation.}
\label{var_fig}
\end{figure}

We compared the performance regarding balancing communication resource allocation for metaverse users under different network bandwidths shown in Fig. \ref{var_fig} (a). We notice that the variance of network resources decreases as the network bandwidth increases for each method. The reason is that with limited resources, the important factors of QoE, such as frame rate and latency, are easily impacted by obtained network resources. Thus, it is hard to maximize overall QoE while reducing imbalance of resource allocation in these conditions. Moreover, SAC-GCN and DQN have lower variances compared to ISAC, GCC-G, and BBR-G in all settings. The reason is that the strategies of SAC-GCN and DQN are centralized learned where the information of other users' states are considered. Moreover, the self-attention mechanism employed by SAC-GCN also helps agents pay more attention to other agents with higher bit rates which further lowers the variance of network resources. 


The performance of all methods in terms of balancing computation resource allocation was studied in Fig. \ref{var_fig} (b). Similar to network resources, the variance of computation resources of these four methods all show a similar downward trend as the available CPU resources increases. In addition, ISAC, GCC-G, and BBR-G always have higher variance since they do not utilize the information from other users when making decisions


\subsection{Resource Utilization Rate Analysis}
\begin{figure}[htbp] 
\centering
\subfigure[]{
\includegraphics[width=0.48\textwidth, height=0.25\textwidth]{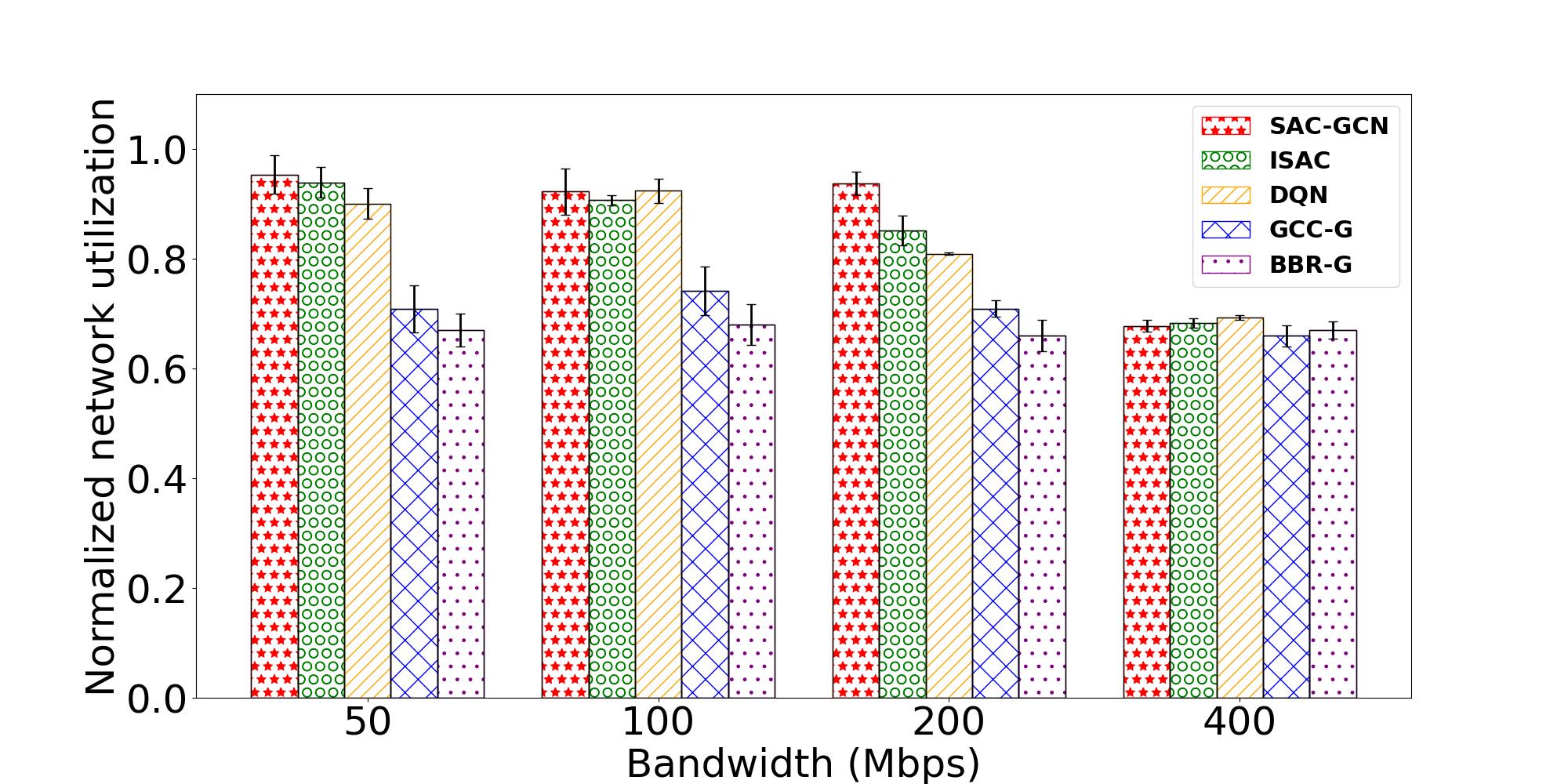}}
\subfigure[]{
\includegraphics[width=0.48\textwidth, height=0.25\textwidth]{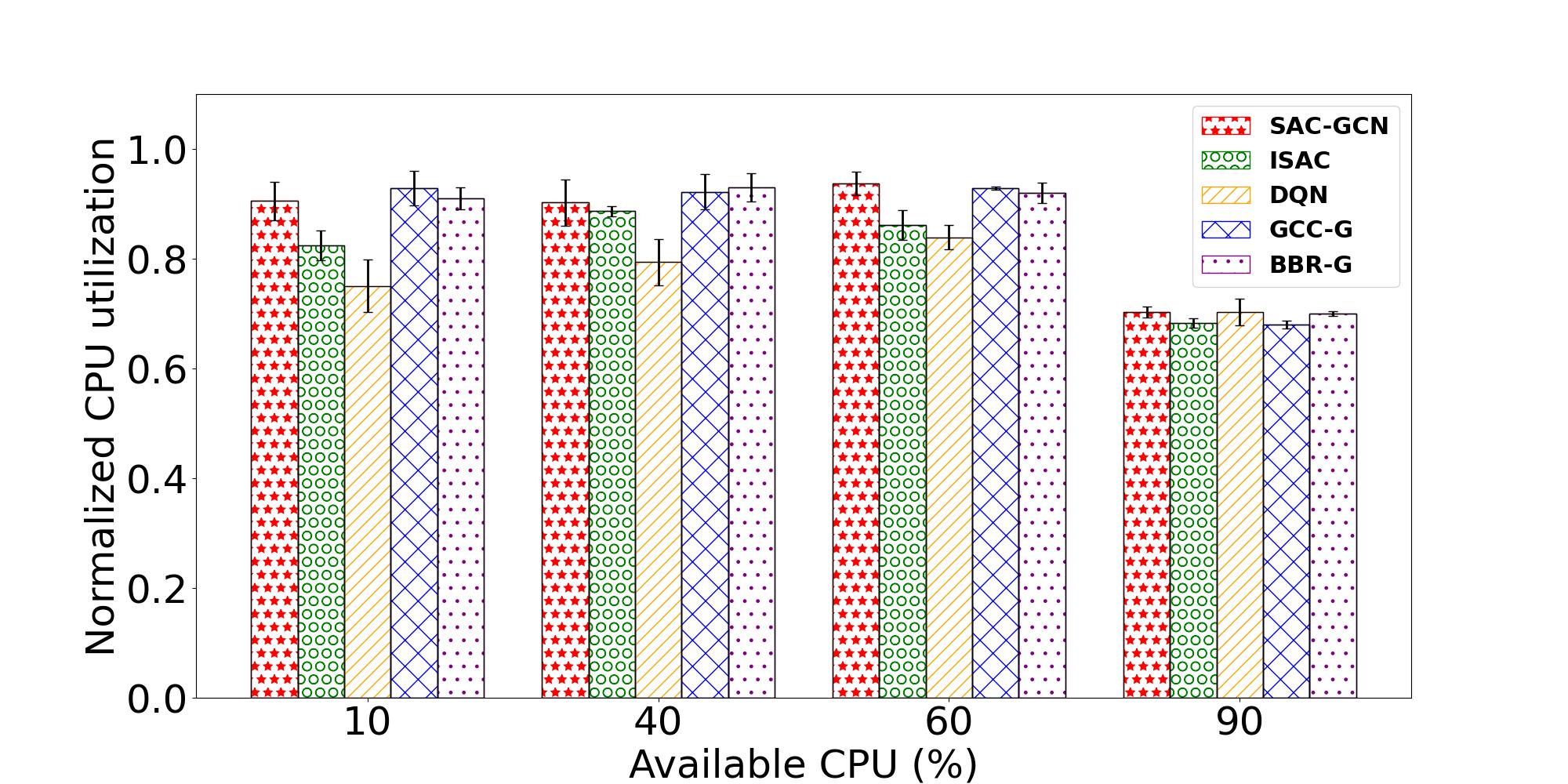}}
\caption{Comparison of the five resource allocation methods regarding resource utilization rate: (a) network resources and (b) CPU resources.}
\label{uti_fig}
\end{figure}

The comparison regarding communication resource utilization rate was conducted under various bandwidths (shown in Fig.\ref{uti_fig} (a)). To ignore the influence of the other resources, we provide sufficient 90\% available CPU resources for each bandwidth. Similar to communication resources, when studying computation resource utilization rate, we also supply sufficient network resources (400Mbps bandwidth) for various available CPU resources (shown in Fig.\ref{uti_fig} (b)).

We observe that there is no clear relationship between resource utilization rate and amount of available resources when resources are insufficient to support all the metaverse users. However, the utilization rates of all methods go down to similar values as more resources are supplied. Since GCC-G and BBR-G employ the greedy algorithm for computation resources, the CPU utilization rates are almost 100\% with insufficient CPU. Although this improves the utilization rate, it may cause other critical problems, i.e., high imbalance of CPU resource allocation. We also observe that our proposed SAC-GCN outperforms other methods most of the time. For other time, it is relatively as good as others regarding network and CPU utilization rate. For example, it reaches 10\%, 15\%, 32\%, and 41\% improvements over ISAC, DQN, GCC-G, and BBR-G for network resource utilization rate, when the network bandwidth is 200Mbps. Moreover, it has 8\% and 12\% improvements for CPU resource utilization rate compared to that of ISAC and DQN when the available CPU is 60\%. This verifies the effectiveness of SAC-GCN for improving resource utilization rates with limited resources.

\subsection{Discussion}
According to the above analysis, we summarize that the DRL-based methods (SAC-GCN, ISAC, and DQN) have much better performance than congestion control methods (GCC-G and BBR-G) in terms of improving overall QoE for users in the metaverse. It is because the trial and error exploration process and the feedback from the environment enable their quick response to the varying multi-user metaverse environment. Thus, these DRL methods can always make better use of the available resources by choosing the appropriate bit rate and corresponding CPU usage percentage for each user. When the resources are insufficient, such as an increase in network delay or the packet loss rate, the change of overall QoE is much more stable compared to that of GCC-G and BBR-G. As for balance of resource allocation, due to the lack of a centrally learned strategy, ISAC cannot perform well in balancing obtained resources for each user, which makes it inappropriate for this problem. On the contrary, our proposed SAC-GCN employs the centralized training and decentralized execution framework where multiple agents are trained with global states in a centralized manner. Moreover, it uses GCN with self-attention mechanism to assist agents to capture more important information in the complex and varying multi-agent environment, which further enhances the cooperation between multiple agents. In addition, our SAC-GCN empowered metaverse resource allocation system can be further improved by incorporating more human-related factors \cite{du2022attention}. For example, eye movement, which can be easily detected by eye trackers on HMDs, is reliable to reflect users’ attention. If the system can adaptively allocate different communication and computation resources to different parts of the metaverse scenes based on users’ attention, users will get the same QoE with less resources consumed.

\section{Conclusion} \label{conclusion}
In this paper, we address the problem of edge resource allocation for multiple users in the metaverse including communication and computation resources. To maximize the trade-off between user experience and balance of resource allocation, we formulate the resource allocation problem as a Dec-POMDP and propose SAC-GCN, a MADRL-based method, where each agent determines the usage of communication and computation resources for one user in the metaverse. To evaluate the performance of SAC-GCN, we design a resource allocation system and carry out many experiments using a virtual city park as a case study. Results demonstrate the effectiveness of SAC-GCN regarding improving overall QoE, balancing resource allocation, and increasing resource utilization rate compared to other methods.


%





\ifCLASSOPTIONcaptionsoff
  \newpage
\fi





\bibliographystyle{IEEEtran}
\bibliography{Bibliography}
%


\vskip -25pt plus -1fil
\begin{IEEEbiography}[{\includegraphics[width=1in,height=1.368in,clip,keepaspectratio]{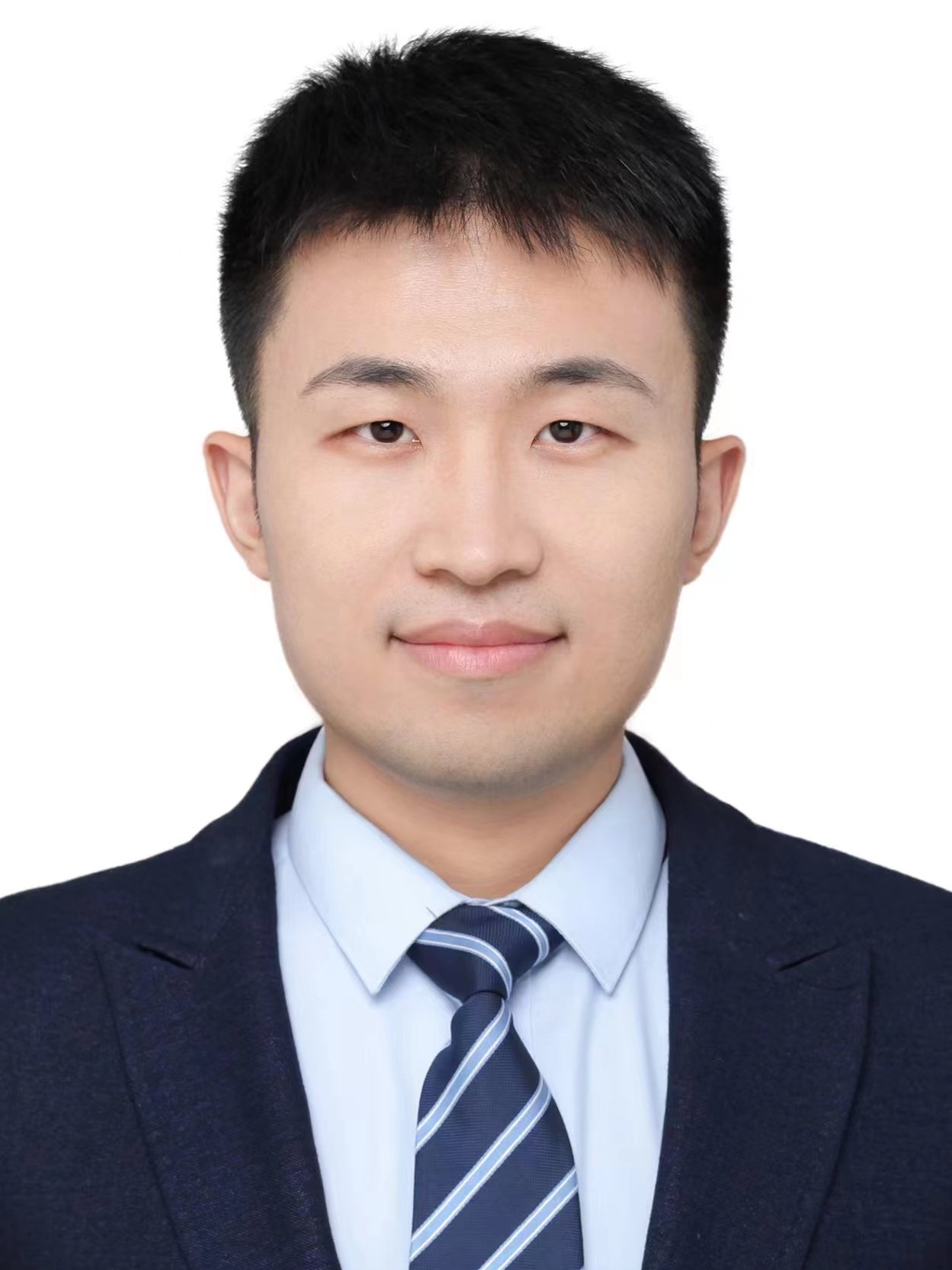}}]{Zijian Long}
(zlong038@uottawa.ca) received the B.Sc. degree in Software Engineering from Beijing Institute of Technology, China, in 2016 and the M.Sc. degree in Electrical and Computer Engineering from the University of Ottawa, Canada, in 2020. He is currently a Ph.D. candidate in the School of Electrical Engineering and Computer Science, University of Ottawa. His research interests include metaverse, XR network, and reinforcement learning.
\end{IEEEbiography}
\vskip -25pt plus -1fil
\begin{IEEEbiography}[{\includegraphics[width=1in,height=1.25in,clip,keepaspectratio]{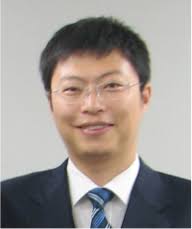}}]{Haiwei Dong}
(haiwei.dong@ieee.org) received the Ph.D. degree from Kobe University, Kobe, Japan in 2010 and the M.Eng. degree from Shanghai Jiao Tong University, Shanghai, China, in 2008. He was a Principal Engineer with Artificial Intelligence Competency Center, Huawei Technologies Canada, Toronto, ON, Canada, a Research Scientist with the University of Ottawa, Ottawa, ON, Canada, a Postdoctoral Fellow with New York University, New York City, NY, USA, a Research Associate with the University of Toronto, Toronto, ON, Canada, and a Research Fellow (PD) with the Japan Society for the Promotion of Science, Tokyo, Japan. He is currently a Principal Researcher with Ottawa Research Center, Huawei Technologies Canada, Ottawa, ON, Canada, and a registered Professional Engineer in Ontario. His research interests include artificial intelligence, multimedia, metaverse, and robotics. He also serves as a Column Editor of IEEE Multimedia Magazine; an Associate Editor of ACM Transactions on Multimedia Computing, Communications, and Applications; and an Associate Editor of IEEE Consumer Electronics Magazine.
\end{IEEEbiography}
\vskip -25pt plus -1fil
\begin{IEEEbiography}[{\includegraphics[width=1in,height=1.25in,clip,keepaspectratio]{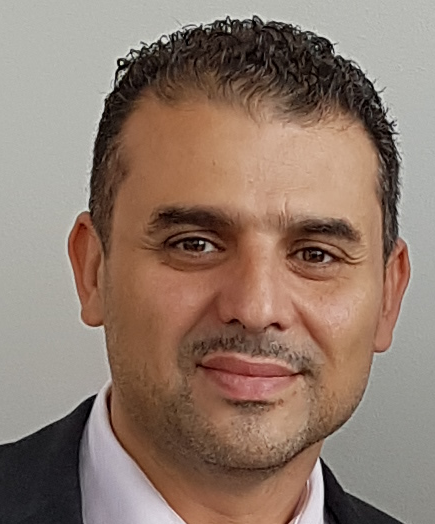}}]{Abdulmotaleb El Saddik}
(elsaddik@uottawa.ca) is currently a Distinguished Professor with the School of Electrical Engineering and Computer Science, University of Ottawa. He has supervised more than 120 researchers. He has coauthored ten books and more than 550 publications and chaired more than 50 conferences and workshops. His research interests include the establishment of digital twins to facilitate the well-being of citizens using AI, the IoT, AR/VR, and 5G to allow people to interact in real time with one another as well as with their smart digital representations. He received research grants and contracts totaling more than \$20 M. He is a Fellow of Royal Society of Canada, a Fellow of IEEE, an ACM Distinguished Scientist, and a Fellow of the Engineering Institute of Canada and the Canadian Academy of Engineers. He received several international awards, such as the IEEE I\&M Technical Achievement Award, the IEEE Canada C.C. Gotlieb (Computer) Medal, and the A.G.L. McNaughton Gold Medal for important contributions to the field of computer engineering and science.
\end{IEEEbiography}





\end{document}